\documentclass[twocolumn,trackchanges]{aastex631}
\newcommand{\kms}{{\rm km\, s}^{-1}}
\newcommand{\muasyr}{\mu{\rm as\, yr}^{-1}}

\shorttitle{LG mass in light of {\it{Gaia}}}
\shortauthors{Benisty et al.}
\graphicspath{{./}{figures/}}

\usepackage{amsmath,amssymb}
\begin{document}

\title{\Large{The Local Group Mass in the light of \textit{Gaia}}}
\newcommand{\afffias}{Frankfurt Institute for Advanced Studies (FIAS), Ruth-Moufang-Strasse~1, 60438 Frankfurt am Main, Germany}
\newcommand{\affbgu}{Physics Department, Ben-Gurion University of the Negev, Beer-Sheva 84105, Israel}
\newcommand{\affcam}{DAMTP, Centre for Mathematical Sciences, University of Cambridge, Wilberforce Road, Cambridge CB3 0WA, UK}
\newcommand{\affcamast}{Institute of Astronomy, University of Cambridge, Madingley Road, Cambridge, CB3 0HA, UK}
\newcommand{\affckavli}{Kavli Institute of Cosmology (KICC), University of Cambridge, Madingley Road, Cambridge, CB3 0HA, UK}
\newcommand{\affmColCam}{Queens’ College, Cambridge, CB3 9ET, UK}
\newcommand{\affTex}{Mitchell Institute for Fundamental Physics and Astronomy, Department of Physics and Astronomy, Texas A\&M University, College Station, TX, USA 77843}

\author[0000-0002-9578-3081]{David Benisty}
\affiliation{\affcam}\affiliation{\affckavli}\affiliation{\affmColCam}
\author[0000-0002-5038-9267]{Eugene Vasiliev}
\affiliation{\affcamast}
\author{N. Wyn Evans}
\affiliation{\affcamast}
\author{Anne-Christine Davis}
\affiliation{\affcam}\affiliation{\affckavli}
\author{Odelia V. Hartl}
\affiliation{\affTex}
\author{Louis E. Strigari}
\affiliation{\affTex}

\begin{abstract}
High accuracy proper motions (PMs) of M31 and other Local Group satellites have now been provided by the {\it Gaia} satellite. We revisit the Timing Argument to compute the total mass $M$ of the Local Group from the orbit of the Milky Way and M31, allowing for the Cosmological Constant. We rectify for a systematic effect caused by the presence of the Large Magellanic Cloud (LMC). The interaction of the LMC with the Milky Way induces a motion towards the LMC. This contribution to the measured velocity of approach of the Milky Way and M31 must be removed. We allow for cosmic bias and scatter by extracting correction factors tailored to the accretion history of the Local Group. The distribution of correction factors is centered around $0.63$ with a scatter $\pm 0.2$, indicating that the Timing Argument significantly overestimates the true mass. Adjusting for all these effects, the estimated mass of the Local Group is $ M  = 3.4^{+1.4}_{-1.1} \times 10^{12} M_{\odot}$ (68\% CL) when  using the M31 tangential velocity $v_{\rm tan}= 82^{+38}_{-35}\,\kms$. Lower tangential velocity models with $v_{\rm tan}= 59^{+42}_{-38}\,\kms$ (derived from the same PM data with a flat prior on the tangential velocity) lead to an estimated mass of $ M  = 3.1^{+1.3}_{-1.0} \times 10^{12} M_{\odot}$ (68\% CL). By making an inventory of the total mass associated with the 4 most substantial LG members (the Milky Way, M31, M33 and the LMC), we estimate the known mass is in the range $3.7^{+0.5}_{-0.5} \times 10^{12} \, M_{\odot}$.
\end{abstract}



\section{Introduction} 

The Timing Argument (TA) is a simple way of working out the mass $M$ of the Local Group (LG). In its earliest manifestation ~\citep{Ka59}, the Milky Way and M31 proto-galaxies are assumed to have a small separation at the time of the Big Bang. They travel away from each other in the Hubble flow~\citep[see][for a general relativistic derivation]{BZ16}. If there is enough mass present, their expansion is reversed. Given an estimate of the age of the universe, together with their present separation and velocity of approach, then the equations of motion can be solved to give the mass of the Local Group. \citet{Ka59} did this assuming the Milky Way and M31 are on an exactly radial orbit. \citet{ELB} showed that the problem retained an analytic solution, even if M31 has some tangential motion. 

Since then, many elaborations of the TA have been proposed, including: (i) the effects of the tidal influence of galaxies outside the Local Group, which is minor~\citep{Ra89}; (ii) the introduction of a Cosmological Constant, which manifests itself as an additional expansion term and increases the LG mass by $\approx 10$ \% ~\citep{Partridge:2013dsa}; (iii) corrections for the effects of hierarchical growth of the two galaxies by comparisons with cosmological simulations~\citep{Kr91,Li:2007eg}, which shows the TA is (mostly) unbiased though suffers from cosmic scatter; (iv) successive refinements of the M31 orbit in view of the improving observational accuracy of M31's tangential motion~\citep{vanderMarel:2012xp,vdM19}.

In this {\it Letter}, we revisit the TA. Our motivations are twofold. First, we wish to exploit the new measurement of the proper motion of M31 provided by the {\it Gaia} satellite~\citep{vdM19,Sa21}. Second, we identify sources of systematic error in applications of the TA which needs correction. Recent work, again driven by data from the {\it Gaia} satellite, has shown that the Large Magellanic Cloud is much more massive than originally envisaged. It is pulling the central parts of the Milky Way towards it~\citep{Er21,Pe21, Ga21}, and so the measured line of sight velocity of M31 needs correction before the TA can be applied. This effect on the TA was noted before by \citet{Penarrubia:2015hqa}, though their analysis method differs from that presented here. Equally important, we include the corrections for the TA-derived mass calibrated on pairs of galaxies extracted from cosmological simulations \citep{Ha21}, which have orbits similar to those of M31 and the Milky Way. In this procedure, it is of crucial importance to ensure that the mock pairs match the LG as closely as possible.

The material is arranged as follows. In \S~\ref{sec:massfromkin}, we estimate the known mass in the LG from stellar/satellite kinematics. \S~\ref{sec:massfromTA} implements our corrections to the TA, while \S~\ref{sec:conc} provides a discussion of the results.

\section{The LG Mass Budget from Kinematics}
\label{sec:massfromkin}

The LG mass can be estimated by modelling the kinematics of tracers (halo stars, satellite galaxies and globular clusters or HI gas) around its prominent members. This assumes that the dark matter is clustered around the major galaxies, and not distributed throughout the LG (as originally envisaged in \citealt{Ka59}).

The advent of {\it Gaia} data has substantially reduced the uncertainty on the virial mass of the Milky Way $M_{\rm MW}$, with most recent measurements satisfying $1.17^{+0.21}_{-0.15} \times 10^{12} M_\odot$~\citep[][ see also \cite{Wa19} and \citet{Fr20} for similar results]{Ca19}. M31 is more massive than the Milky Way with dynamical arguments suggesting $M_{\rm M31}= 1.8 \pm 0.5$~\citep[][see also \cite{Diaz:2014kqa} and \cite{KK14} for similar values]{Sh14}. This suggests that the mass associated with the two largest galaxies in the LG is $3.0^{+0.5}_{-0.5} \times 10^{12} M_\odot$. Note that the virial mass of M31 is more uncertain than that of the Milky Way, and it remains (just) possible that the mass ratio is close to unity~\citep[e.g.,][]{Ev00,Fa13,Ka18}, which gives us a lower bound. 


The next most massive members of the LG in decreasing order are: M33 with $ M_{\rm M33} = 5.0\pm1.0 \times 10^{11} M_\odot $ \citep{Co14,Ka17} and the Large Magellanic Cloud with $M_{\rm LMC} = 1.8\pm0.4\times 10^{11} M_\odot$ \citep{Er19, Sh21}. M32 is a compact dwarf elliptical with current mass at least an order of magnitude less than M33. Its progenitor may once have been more massive than M33, though much of its tidally stripped material is now in the halo of M31~\citep{DS}, and so already accounted for in our inventory. We surmise that the minor members of the LG contribute about $0.7 \times 10^{12} M_\odot$ to the mass budget. We conclude that the total mass in the LG -- as judged from kinematics of tracers -- is at least $3.0 \times 10^{12} M_\odot$ and most likely in the range $3.7 \pm 0.5 \times 10^{12} M_\odot$.

\section{The Timing Argument Revisited}
\label{sec:massfromTA}

\subsection{The Data}
\label{subsec:data}

We take the current values of the separation between the Milky Way and M31 as $r = 770 \pm 40$~kpc and the heliocentric line-of-sight velocity as $v_{\rm los} = -301 \pm 1 \, \kms$ \citep{vanderMarel:2012xp}.
The measurement of the proper motion (PM) of M31 has been refined over the last decade. \cite{vanderMarel:2012xp} used {\it Hubble Space Telescope} (HST) observations in three small off-centered fields in conjunction with a model for its internal kinematics, deriving the mean PM of M31 to be $\mu_\alpha=45\pm13\,\muasyr$, $\mu_\delta=-32\pm12\,\muasyr$. This value is dominated by the Solar velocity with respect to the Milky Way centre, which corresponds to $\mu_{\{\alpha,\delta\}}^{\rm reflex}=\{38,\; -22\}\,\muasyr$ at the distance of M31 ($770\pm 40$~kpc). Thus the reflex-corrected PM is consistent with zero, and they gave a $1\sigma$ upper limit on the tangential velocity at $34\,\kms$. More recently, \citet{vdM19} computed an independent estimate of M31's absolute PM from \textit{Gaia} Data Release 2, which had twice larger uncertainties than the \textit{HST}-based value, and is also larger in absolute sense: $\mu_{\alpha,\delta}=\{65\pm18,\; -57\pm15\}\,\muasyr$, with an additional systematic uncertainty of $16\,\muasyr$ in each component. The error-weighted average of the two independent measurements is $\mu_{\alpha,\delta}=\{49\pm11,\; -38\pm11\}\,\muasyr$ and corresponds to a reflex-corrected tangential velocity of $57^{+35}_{-31}\,\kms$. 
\citet{Sa21} used the updated \textit{Gaia} Early Data Release 3 astrometry to measure $\mu_{\alpha,\delta}=\{49\pm11,\; -37\pm8\}\,\muasyr$, which is very close to the weighted average derived in \citet{vdM19}; however, they reported the reflex-corrected tangential velocity to be $82\pm 31\,\kms$. The discrepancy between the two studies stems from the way the distribution of tangential velocities is derived from the distribution of PM.
\citet{vdM19}, following their earlier work (section 3.1 in \citealt{vdMG08}), derive the posterior distribution of the magnitude of the two-dimensional tangential velocity vector $\boldsymbol v_\mathrm{tan}$ by convolving the observed Gaussian PM distribution with a prior on the tangential velocity $\mathcal P\big( |\boldsymbol v_\mathrm{tan}| \big)$, which they take to be flat in $|\boldsymbol v_\mathrm{tan}|$, favouring smaller values. 
By contrast, \citet{Sa21} simply convert both reflex-corrected PM components into velocity and sum them in quadrature, which effectively means using a flat prior on each component of $\boldsymbol v_\mathrm{tan}$, i.e., $\mathcal P\big( |\boldsymbol v_\mathrm{tan}| \big) \propto |\boldsymbol v_\mathrm{tan}|$.

As there is no convincing reason in favour or against the use of \citet{vdMG08}'s prior, we consider both alternatives, drawing Monte Carlo samples from the distribution of observed PM values (taken from \citealt{Sa21} in both cases) with or without the additional reweighting by the prior. It is intuitively clear that increasing the M31 tangential velocity makes the M31 orbit less eccentric and therefore the LG mass must increase (for fixed age of the Universe).
\begin{figure}
 	\centering
\includegraphics[width=0.45\textwidth]{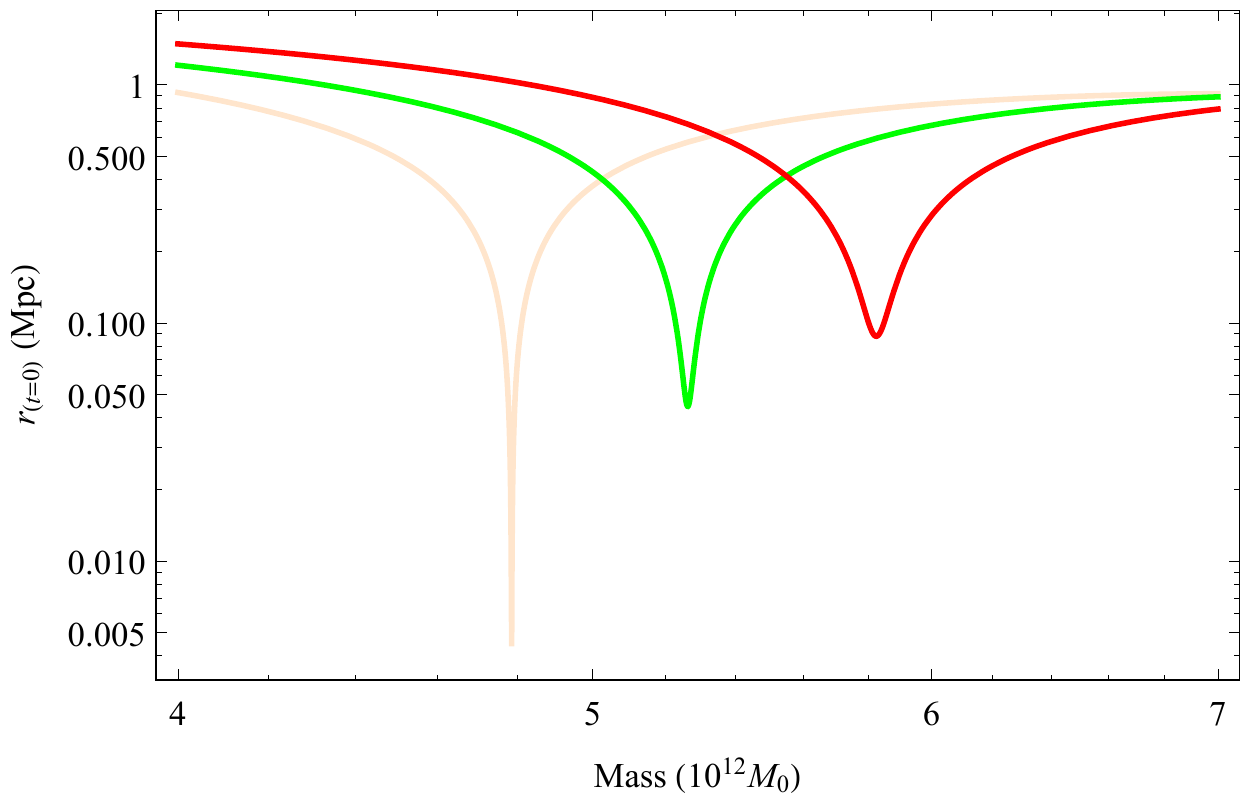}
\caption{The separation of the LG as $t \rightarrow 0$ (the Big Bang) for different masses with different tangential velocities:  $v_{\rm tan} = 17 \, \kms$ ~\citep[orange,][]{vanderMarel:2012xp}; $v_{\rm tan} = 57 \, \kms$~\citep[green,][]{vdM19} and $v_{\rm tan} = 82.5 \, \kms$~\citep[red,][]{Sa21}. The minimum separation gives the mass of the LG implied by the TA.} 
 	\label{fig:Mvsvt}
\end{figure}

\subsection{The TA Algorithm with Cosmological Constant}

\label{sec:ETBP}
The center of mass coordinate system is defined by the relative distance $r = |\vec{r}_{\rm M31} - \vec{r}_{\rm MW}|$ and the relative velocity $\vec{v} = d\vec{r}/dt$. The masses are replaced by the total mass $M := m_{\rm MW} + m_{\rm M31}$. In polar coordinates $(r,\varphi)$, the relative distance variation now reads \citep{Emelyanov:2015ina,Carrera:2006im,2013MNRAS.429.3477E}:
\begin{equation}\label{ENL}
\ddot{r} = \frac{l^2}{r^3}-\frac{GM}{r^2} +  \frac{1}{3}\Lambda c^2 \, r,
\end{equation}
where $l$ is the conserved angular momentum per mass 
($l=r^2 \dot\varphi = r \, v_{\rm{tan}}$). Based on \cite{Partridge:2013dsa} we include the cosmological constant: $\Lambda = (4.24 \pm 0.11) \times 10^{-66} \, {\rm eV}^2$ as determined by the latest Planck measurements \citep{Planck:2018vyg}. We integrate orbits back in time to the Big Bang using the Age of the Universe $t_0 = 13.799\pm 0.021$ Gyrs, also taken from the Planck measurements.

Note that the inclusion of the Cosmological Constant means that the mass of the LG inferred from the TA is no longer analytic. To calculate it, we reverse the direction of time and compute the separation at the Big Bang. When the curve is at a minimum, this is the mass implied by the TA. Fig~\ref{fig:Mvsvt} illustrates the method by showing the separation at the Big Bang against mass of LG for different $v_{\rm tan}$. As expected, an increase in the tangential velocity implies a larger $M$ value.

\begin{figure}
\includegraphics{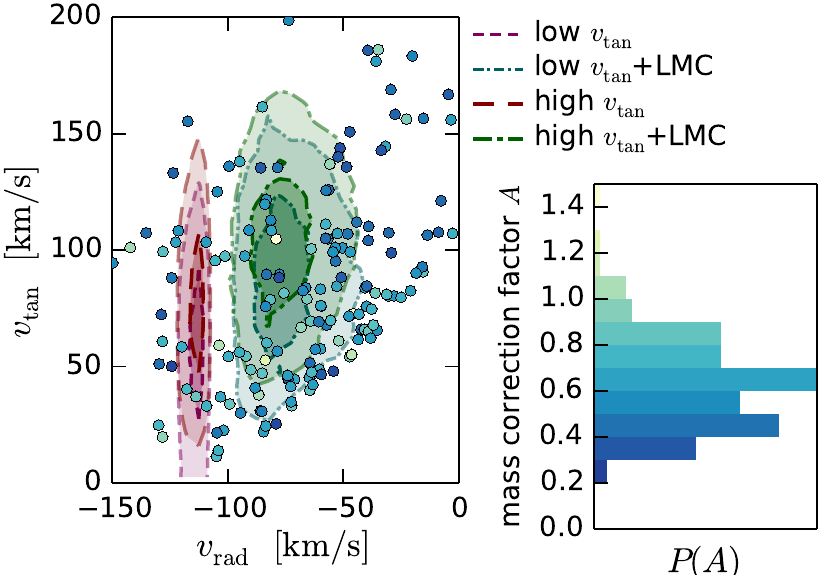}
\caption{Posterior distribution of values of M31's radial and tangential velocity for two sets of Monte Carlo samples generated from the same PM measurements by \citet{Sa21}: using the prior on the magnitude of $v_\mathrm{tan}$ from \citet{vdMG08} results in lower values (short dashes), while using the raw PM measurements without any reweighting produces higher values (long dashes). In both cases, the radial velocity is $\simeq -114\pm1\,\kms$. After compensating for the LMC perturbation as described in \S~\ref{sec:LMC}, we find that $v_\mathrm{tan}$ is increased by 25--30\,$\kms$, and $v_\mathrm{rad}$ is shifted to $-75\pm15\,\kms$. This distribution is shown by short and long dot-dashed contours and shaded in green. These distributions are compared to the radial and tangential velocities of galaxy pairs selected from the IllustrisTNG cosmological simulation, in which the mass correction factor is computed as the ratio of the actual combined mass of both galaxies to the mass obtained from TA, as described in \citet{Ha21} and \S~\ref{sec:CS}. Points in the left panel are coloured according to the correction factor, while the right panel shows the histogram of correction factors in the sample of galaxy pairs.}  
\label{fig:vrad_vtan_distribution}
\end{figure}

\begin{figure*}
 	\centering
\includegraphics{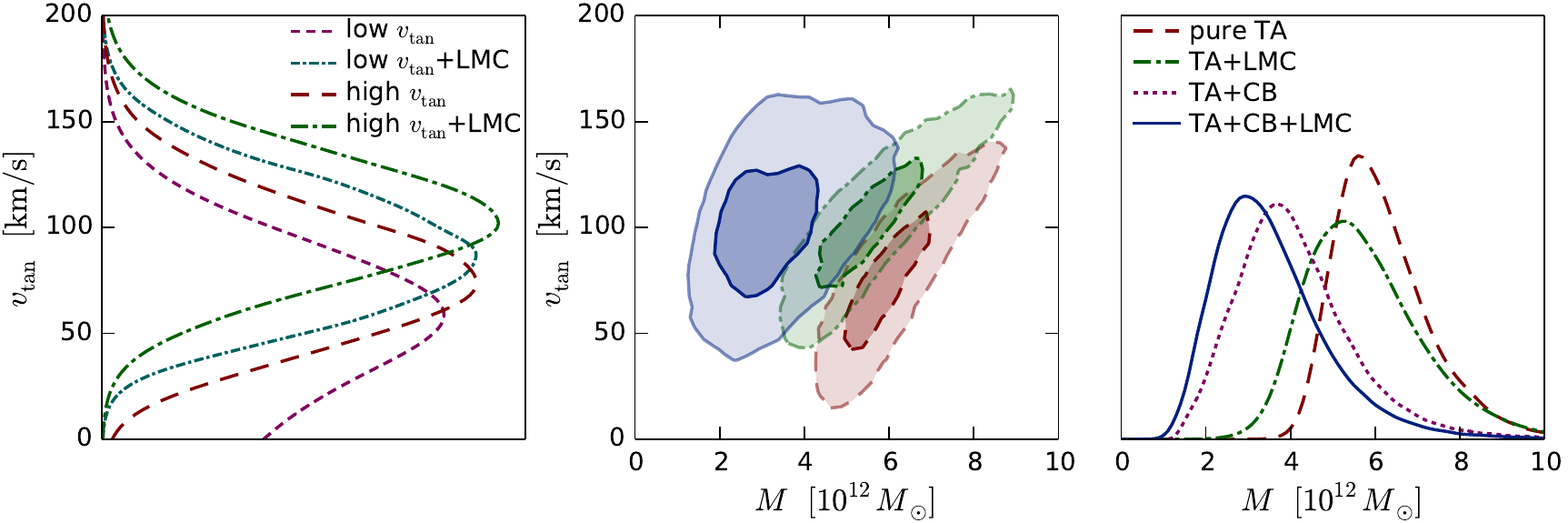}
 \caption{The posterior distribution of the tangential velocity and the inferred LG mass. \textbf{Left panel} shows the distribution of M31's tangential velocities in two cases: ``low'' (short dashes) uses the prior from \citet{vdMG08}, ``high'' (long dashes) uses just raw proper motion measurements from \citet{Sa21} without any further reweighting. In both cases we also show the velocity that would be measured in absence of the LMC, as explained in \S~\ref{sec:LMC} (short and long dot-dashes respectively). \textbf{Right panel} shows the distribution of LG masses for the high $v_\mathrm{tan}$ case only (the results for the low $v_\mathrm{tan}$ case are very similar), in four possible combinations: pure TA (dashed red), compensating the LMC perturbation (dot-dashed green), multiplying the inferred mass by a cosmic bias correction factor $A$ sampled from the distribution obtained by \citet{Ha21} as explained in \S~\ref{sec:CS} (dotted purple), and including both corrections (solid blue). \textbf{Centre panel} shows the $1\sigma$ and 2$\sigma$ contours of the  two-dimensional posterior distribution of both quantities for three of these cases (omitting TA+CB for clarity), using the same colours and line styles as in the right panel. 
 } 
  	\label{fig:MvsvtDis}
\end{figure*} 

\subsection{The Effect of the Large Magellanic Cloud}  \label{sec:LMC}

The Milky Way is peculiar in having an unusually large, nearby satellite galaxy, the Large Magellanic Cloud (LMC). The importance of this interloper for the Milky Way has become clear over the last years~\citep{2015ApJ...802..128G,Er21,Pe21,Ga21}. The central part of the Milky Way (which includes the Solar neighbourhood) is pulled downwards towards the LMC on its pericentre passage, though the sluggish outer parts stay put. The measured heliocentric line-of-sight velocity and PM of M31 therefore include a contribution due to this downward motion, which we would like to remove.

\citet{Penarrubia:2015hqa} accounted for the presence of the LMC by assuming that it forms a two-point-mass system with the Milky Way, and that M31 moves around the barycentre of this combined system. The displacement and velocity shift of the Milky Way relative to the barycentre are obtained by multiplying the relative position and velocity of the LMC in the Milky Way-centered frame by the mass ratio $M_\mathrm{LMC} / (M_\mathrm{MW}+M_\mathrm{LMC})$. These offsets need to be subtracted from the current position and velocity of M31 in the Milky Way-centered frame prior to computing its trajectory in the barycentric system. This argument is qualitatively correct, but ignores the fact that the LMC is currently only $\sim50$~kpc from the Milky Way centre, and that the enclosed mass of the Milky Way within this radius is substantially smaller than its total mass -- in other words, it underestimates the actual displacement of the central region of the Milky Way caused by the LMC.

A more sophisticated technique to compensate for the LMC perturbation was recently introduced by \citet{CM21}. It starts by computing the past trajectory of the Milky Way and the LMC under their mutual gravitational attraction, using the actual (distance-dependent) force from each galaxy rather than the point-mass approximation implied by \citet{Penarrubia:2015hqa}'s method. Once this has been done for a given choice of Milky Way and LMC potentials, we integrate the orbit of M31 in this time-dependent potential of both galaxies backward in time until the LMC perturbation is negligible, and the integrate it forward without the LMC to the present epoch. For simplicity, the Milky Way potential is fixed to an NFW halo with virial mass $M_{\rm vir} =1.1 \times 10^{12} M_\odot$, virial radius $r_{\rm vir}=270$ kpc and a concentration $c=13.5$, but we take into account the uncertainty on the LMC mass by sampling it from a log-normal distribution centered on $\log_{10}( M_{\rm LMC} )=11.15$ with width 0.15 dex and repeating the orbit rewinding step for each choice of LMC mass. 

The left-hand panel of Figure~\ref{fig:vrad_vtan_distribution} shows the posterior distribution of M31's Galactocentric radial and tangential velocity components $v_\mathrm{rad}$, $v_\mathrm{tan}$, with or without the LMC correction. The use of a prior from \citet{vdMG08} results in a lower tangential velocity, $v_\mathrm{tan} = 59 \pm 34\,\kms$, while the use of raw PM measurements produces a higher $v_\mathrm{tan} = 78 \pm 32\,\kms$. In both cases, the compensation of the LMC perturbation increases $v_\mathrm{tan}$ by 25--30~$\kms$, changes $v_\mathrm{rad}$ from $-114\pm 1\,\kms$ to $-75\pm 15\,\kms$, and increases the distance by $\sim 40$~kpc. The marginalized posterior distributions of $v_\mathrm{tan}$ for all four cases are also shown in the left-hand panel of Figure~\ref{fig:MvsvtDis}. Compared to the simpler model for the barycentric motion used in \citet{Penarrubia:2015hqa}, our velocity correction is roughly twice higher for the given LMC mass, but since that paper used a significantly larger range of LMC masses with a median at $2.5\times 10^{11}\,M_\odot$, the velocity correction is quite similar in absolute terms.

\subsection{Cosmic Bias and Scatter}  \label{sec:CS}


Owing to simplifications in the TA, the mass estimate may suffer from systematic bias and scatter. \citet[][see their Figure 1]{Li:2007eg} found that the TA mass is unbiased, though with some scatter, using analogues of the Milky Way--M31 pair extracted from the dissipationless Millenium simulation. However, \citet{Go14} noted the TA mass is only unbiased on average, and can be an overestimate if the pairs are restricted to 
to have similar radial and tangential velocities as the true Milky Way and M31. The matter has been re-investigated recently by \citet{Ha21}, who used the IllustrisTNG N-body and hydrodynamical simulations. They also found a tendency of the TA mass to be overestimated. Specifically, \citet{Ha21} identify 580 bound analogues of the LG by a series of cuts on $B$ band magnitude, separation, velocity of approach and total velocity, computing distributions of $P(A)$, where $A$ is the ratio of true mass to mass predicted by the TA. 
We tailor the \citet{Ha21} sample by imposing three new cuts: (i) a separation between 650 and 950 kpc;
(ii) a mass ratio within a factor of 4;
(iii) $-150 < v_{\rm rad} < v_{\rm tan}-100\,\kms$ so
it resembles the actual distribution of LMC-corrected velocities as shown in Figure~\ref{fig:vrad_vtan_distribution}. This retains 160 galaxy pairs, with the distribution $P(A)$ shown the right panel of that figure centered around $A=0.63$ with a scatter $\pm0.2$. 

To incorporate uncertainties, we use the Markov chain Monte Carlo method. We sample $10^4$ values for the initial conditions vector $\{ {\tilde r}, {\tilde v}_{\rm rad}, {\tilde v}_{\rm tan}, t_0, \Lambda \}$ and calculate the corresponding predicted mass, and then convolve this TA-predicted mass distribution with the distribution of correction factors $P(A)$. We assume the initial conditions for $t_0$ and $\Lambda$ have a normal distribution with the mean and dispersion given by the estimate from Planck and its reported uncertainty.

\begin{table}
 	\centering
 	\hspace*{-1.5cm}
\begin{tabular}{| c | c c c| }
\hline\hline
Model & $M\,(10^{12} M_{\odot})$ && $P_1$ \\
\hline\hline
pure TA & $6.0^{+1.3}_{-0.9}$ && 0.008 \\
TA+LMC & $5.6^{+1.6}_{-1.2}$ && 0.10 \\
TA+CB & $3.9^{+1.5}_{-1.1}$ && 0.32 \\
TA+CB+LMC & $3.4^{+1.4}_{-1.1}$ && 0.29 \\
same, low $v_\mathrm{tan}$ & $3.1^{+1.3}_{-1.0}$ && 0.26 \\
\hline
 \end{tabular}
 \caption{The predicted TA mass under different assumptions on the M31's tangential velocity and including or not the correction for the LMC and cosmic bias (CB) separately and together. First four lines use a flat prior resulting in higher values for $v_\mathrm{tan}$, the last one uses a prior from \citet{vdMG08} resulting in lower $v_\mathrm{tan}$, which reduces the inferred LG mass by $\lesssim 10\%$. $P_1$ is the posterior probability enclosed by the observational range $3.7^{+0.5}_{-0.5}\times 10^{12} M_\odot$. }
  	\label{table:tabone}
\end{table} 
\begin{figure}
 	\centering
\includegraphics[width=0.45\textwidth]{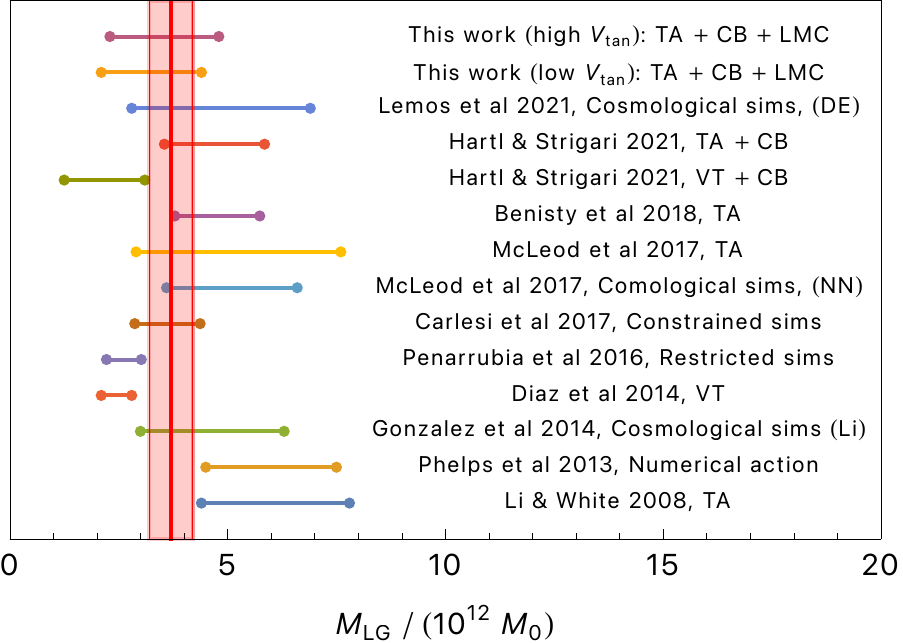}
 \caption{Comparison of recent estimates of the LG mass, shown as best fit and 68\% confidence intervals. The red vertical band shows the range $3.7^{+0.5}_{-0.5}\times 10^{12} M_\odot$ computed using the recent estimates of the mass of the Milky Way and M31 with the claimed
 observational errors. (Here, VT = Virial Theorem, TA = Timing Argument, CB = Cosmic Bias. A number of authors used LG analogues extracted from cosmological simulations, but differed in use of interpolating methods, namely NN = Neural Networks, Li = Likelihood, DE = likelihood-free density estimation.) }
  	\label{fig:final}
\end{figure} 

\section{Results and Discussion}
\label{sec:conc}

The right panel of Fig.~\ref{fig:MvsvtDis} shows the posterior distributions of the the total LG mass obtained for the high M31 tangential velocity case, including or not the correction for the LMC perturbation and cosmic scatter. The central panel shows the two-dimensional posterior distributions. The results are summarized in Table~\ref{table:tabone}, together with the low tangential velocity case. We also record the posterior probability $P_1$ that $M$ lies in the range $3.7^{+0.5}_{-0.5}\times 10^{12} M_\odot$. This is informed by our inventory of the LG mass in \S~2.

The effect of correcting for the LMC's perturbation is to decrease the inferred LG mass by $\sim 10 \%$. Although the correction increases the tangential velocity of M31, which normally would lead to a higher LG mass from TA, it also affects the radial velocity and the distance, so the net result is the opposite (a downward shift). To lowest order, this counteracts the effect of the Cosmological Constant, which acts in the opposite sense by a similar amount~\citep[e.g.,][]{Partridge:2013dsa,Be19,Be20}. Note that further corrections to the infall velocity are probably also needed because of the effects of M33 and M32. M33 and M31 came within $\sim 50$ kpc of each other in the past, $\sim 6.5$ Gyr ago~\citep{TG20}, whilst M32 may even have been more massive than M33 before its catastrophic encounter with M31. Although the effects of these interactions require detailed modelling, they are probably $\lesssim 10\%$ (as less important than the LMC).

The effect of correction for cosmic bias and scatter is substantial, reducing the median LG mass by a factor $\sim 1.5$ and increasing its relative uncertainty by a similar amount. In constructing our distribution of correction factors $A$, we ensured as much as possible that our mock LGs match the distribution of LMC-corrected velocities of infall. There is a wide range of accretion histories in any mock LGs extracted from simulations. It is important to condition distributions on the true environment of the LG as much as possible, as first clearly realised by \citet{Go14}. The pure TA mass -- in our case, $6.0^{+1.3}_{-0.9} M_\odot$ -- can then be a serious overestimate of the true mass. From Table~\ref{table:tabone}, we see that the mass of the LG is $3.4^{+1.4}_{-1.1} \times 10^{12} M_\odot$ with the raw data (the high $v_{\rm tan}$ case). If the \citet{vdMG08} prior is used, then the mass is $3.1^{+1.3}_{-1.0} \times 10^{12} M_\odot$ (the low $v_{\rm tan}$ case). Both are in reasonable accord with the dynamically estimated LG mass as quoted in \S~\ref{sec:massfromkin}. 

Since the LG is assumed to be a closed system, the total energy is negative $E < 0$, otherwise M31 could approach infinity. From this limit, we get an expression for the minimal mass \citep[cf][]{Ch09}: $G M_{\rm min} = r v^2/2 - \Lambda c^2  r^3/6$. For the low tangential velocity, the minimal mass is $1.27 \pm 0.11 \times 10^{12} M_{\odot}$, and for the high tangential velocity, the minimal mass is a bit larger: $1.61 \pm 0.24 \times 10^{12} M_{\odot}$. These numbers may be compared with the observationally derived lower limit to the LG mass of $\approx 3 \times 10^{12}M_\odot$ in \S~2.

Fig.~\ref{fig:final} compares the value obtained in this paper with other recent measurements. Notice that our masses are somewhat lower than (though still consistent with) a number of other recent estimates, such as numerical implementation of least action~\citep{Ph13,Banik:2016emv}, simulations~\citep{Go14,Ca17}, neural networks~\citep{Mc17} and likelihood-free density estimation~\citep{Lemos21}. However, estimates in Fig.~\ref{fig:final} based on the virial theorem \citep[e.g.,][]{Diaz:2014kqa,Ha21} or on the assumption of pure radial orbits~\citep{Penarrubia:2015hqa} are systematically lower than our values. Whilst the outer parts of the Local Group are not virialized, \citet{Ha21} showed that that virial mass estimator is unbiased, but the scatter around the true value is much larger for virial mass estimators than for the TA, rendering it a much less satisfactory method. The high radial velocities of some outlying satellites like NGC 3109 have been suggested as evidence for a past encounter between the Milky Way and M31~\citep{Banik:2016emv}. This is possible in some modified gravity theories, though not in the $\Lambda$CDM paradigm used in this paper.

Overall, we conclude that the LG mass derived via the TA -- if calibrated against realistic analogues in simulations  -- is in reasonable agreement with the mass known to be associated with the Milky Way, the LMC, M31 and M33. {\it Gaia} has improved the accuracy with which the first two are known. The main observational uncertainty that remains is the virial mass of M31, on which future work could usefully be concentrated. But, the Timing Argument works better than we have a right to expect such a simple argument to do -- once corrected from the effects of the LMC and Cosmic Bias! 

\begin{acknowledgements}
EV is supported by the Consolidated Grant, whilst DB thanks the Blavatnik and the Rothschild Foundations for support and partial support from European COST actions CA15117 and CA18108.
\end{acknowledgements}

\bibliography{ref}
\bibliographystyle{aasjournal}

\end{document}